\documentstyle[psfig]{mn}

\title[The `double-double' source B\,1834+620]{The radio source B\,1834+620: A double-double radio galaxy with interesting properties}
\author[A.P. Schoenmakers et al.]{Arno P. Schoenmakers$^{1,2,3}$\thanks{Current address: NFRA, P.O. Box 2,
    7990 AA, Dwingeloo, The Netherlands; Email: Schoenmakers@nfra.nl}, A.G. de Bruyn$^{3,4}$, H.J.A. R\"{o}ttgering$^2$, H. van der Laan$^1$\\ 
$^1$ Astronomical Institute, Utrecht University, P.O. Box 80\,000, 3508~TA Utrecht, The Netherlands\\
$^2$ Sterrewacht Leiden, Leiden University, P.O. Box 9513, 2300~RA Leiden, The 
Netherlands\\
$^3$ NFRA, P.O. Box 2, 7990~AA Dwingeloo, The Netherlands\\
$^4$ Kapteyn Astronomical Institute, University of Groningen, P.O. Box 800, 9700~AV Groningen, The Netherlands}
\date{Received ; accepted}
\pubyear{1998}

\begin{document}
\maketitle

\begin{abstract}
We present a study of the peculiar radio galaxy B\,1834+620. It is
characterised by the presence of a 420-kpc large edge-brightened radio
source which is situated within, and well aligned with, a larger (1.66
Mpc) radio source. Both sources apparently originate in the same host
galaxy, which has a ${\rm R_s}$-magnitude of 19.7 and a redshift of
0.5194, as determined from the strong emission-lines in the spectrum.
We have determined the rotation measures towards this source, as well
as the radio spectral energy distribution of its components.  The
radio spectrum of the large outer source is steeper than that of the
smaller inner source. The radio core has a spectrum that peaks at a
frequency of a few GHz.  The rotation measures towards the four main
components are quite similar, within $\sim\!2$ rad m$^{-2}$ of 58 rad
m$^{-2}$.  They are probably largely galactic in origin. We have used
the presence of a bright hotspot in the northern outer lobe to
constrain the advance velocity of the inner radio lobes to the range
between $0.19c$ and $0.29c$, depending on the orientation of the
source. This corresponds to an age of this structure in the range
between 2.6 and 5.8 Myr. We estimate a density of the ambient medium
of the inner lobes of $\la 1.6\times10^{-30}$ gr\,cm$^{-3}$ (particle
density $\la8\times10^{-7}$ cm$^{-3}$).  A low ambient density is
further supported by the discrepancy between the large optical
emission-line luminosity of the host galaxy and the relatively low
radio power of the inner lobes.

\end{abstract}


\section{Introduction}

Powerful radio galaxies of type FRII (Fanaroff \& Riley 1974) are characterized by extended radio lobes with
leading compact, bright features, the so-called `hotspots', and often a
compact central radio core. 
It is almost certain now that the extended lobes are formed by two
relativistic jets emerging from a supermassive black hole at the center of
a massive galaxy (e.g. Scheuer 1974; Blandford \& Rees 1974). The large
size that these radio sources can obtain indicates that the jet formation
proces must be a long-lived (though not necessarily continuous)
phenomenon. Estimated ages for extended radio sources reach up to $10^8$
yr (e.g. Alexander \& Leahy 1987). Also, the elongated structure of
FRII-type radio galaxies and the relative straightness of the sometimes
observed radio jet indicates that the direction of the outflow is
relatively stable during the activity time.

Schoenmakers et al. (1999, hereafter Paper I) have presented a small
number of sources which consist of two pairs of double-lobed radio
sources which are well aligned and apparently hosted by the same
galaxy.  Because of their peculiar radio morphology,
they proposed to call these `double-double' radio galaxies (DDRGs).
Schoenmakers et al. conclude that the most likely mechanism to explain
these properties is a interruption of the central jet producing
mechanism. Kaiser, Schoenmakers \& R\"{o}ttgering (1999, hereafter
Paper II) have elaborated such a restarted jet scenario. They find
that the density in the `old' cocoons hosting the outer radio lobes is
too low to explain the currently observed inner radio structures, and
they conclude that some material from the outside must have penetrated
the `old' cocoon and thus increased its density. Kaiser et al. (1999)
find that warm ($10^4$ K) clouds embedded in the hot ($10^8$ K)
InterGalactic Medium (IGM) are probably able to survive the passage of
the bowshock trailing the expansion of the `old' cocoon and will
penetrate into the cocoon, where they will be shredded on timescales
of order $10^7$ yr. This may increase the density of the cocoon to
the values necessary to explain the radio properties of the inner
structures.
 
The radio source B\,1834+620 is one of the sources first presented in
Paper I and, since it was the first DDRG to be discovered by us, it
has been the subject of several follow-up studies. In this paper we
report on our radio and optical observations of it. In
Sect. \ref{sec:data} we present our radio and optical data.  Section
\ref{sec:results} presents the main results obtained from these
observations.  In Sect. \ref{sec:discussion} we estimate the age of the
inner source components and derive the density of the medium
surrounding the inner structure.  Our conclusions are presented in
Sect. \ref{sec:conclusions}.

We use $H_0 = 50$\,km\,s$^{-1}$\,Mpc$^{-1}$ and $q_0=0.5$ throughout
this paper. A radio spectral index $\alpha$ is defined as $S_{\nu}
\propto \nu^{\alpha}$. The radio contour maps presented in this paper
have been rotated by 30 degrees in a counter-clockwise (CCW)
direction, unless indicated otherwise.

\begin{table}
\setlength{\tabcolsep}{4pt}
\caption{\label{tab:1834_radio} Log of the radio observations of
B\,1834+620.}
\begin{tabular}{l l l l r@{$\times$}l l}
\hline\\
Freq. & \multicolumn{1}{c}{Date} & \multicolumn{1}{c}{Instr.} & \multicolumn{1}{c}{Survey} & \multicolumn{2}{c}{Beam} & rms \\
$[{\rm MHz}]$ & & & & $[\arcsec$ & $\arcsec\,]$ & $[{\rm mJy}]$ \\
\hline \\
325 & Jan. 1991   & WSRT    & WENSS & 54.0 & 61.1 & 3.0  \\
612 & Dec. 1990   & WSRT    & WENSS & 28.0 & 31.7 & 3.0  \\
840 & Mar. 1998   & WSRT    &       & 20.7 & 23.9 & 0.4  \\
1395 & Feb. 1995  & WSRT    &       & 27.7 & 10.1 & 0.3  \\ 
1400 & Apr. 1995  & VLA-D   & NVSS  & 45.0 & 45.0 & 0.45 \\
1435 & Jan. 1997  & VLA-A   &       & 1.49 & 1.03 & 0.07 \\
4850 & Jun. 1998  & VLA-BnA &       & 1.00 & 0.70 & 0.06 \\
8460 & Aug. 1996  & VLA-D   &       & 8.44 & 6.17 & 0.06 \\
\hline \\
\end{tabular}
\end{table}

\section{Observations}
\label{sec:data}
In this section we present radio and optical observations of B\,1834+620. Details of the radio observations presented here can be found in Tab. \ref{tab:1834_radio}.

\begin{figure}
\psfig{figure=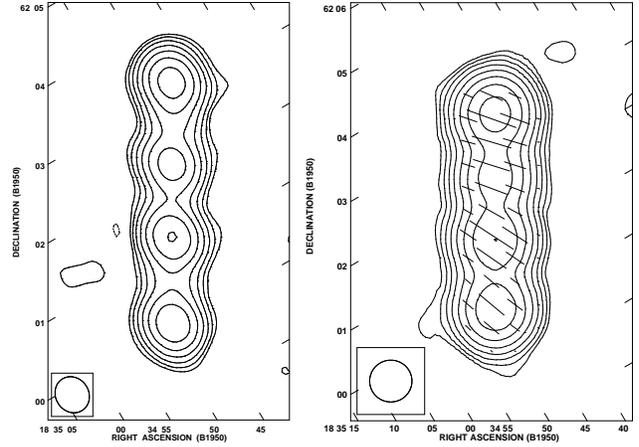,angle=0,width=\columnwidth} 
\caption{\label{fig:1834_wenss}Contour plots of B\,1834+620 from the
WENSS survey at 612 MHz (a, left) and from the NVSS survey at 1.4 GHz
(b, right). On the WENSS map the contours start at 5 mJy beam$^{-1}$, on the
NVSS map at 1.3 mJy beam$^{-1}$. Subsequent contours are logarithmic
with a factor of 2 between levels. Dashed contours designate negative
levels. Also plotted are the observed vectors of the $E$-field
of the linearly polarized intensity at 1.4 GHz. Their length
corresponds to the polarized intensity and is 0.5 mJy beam$^{-1}$ per
arcsec.}
\end{figure}

\subsection{WENSS and NVSS data}

The WENSS survey (Rengelink et al. 1997) has mapped the sky northwards
of $+28\degr$ Dec. at 325 MHz. About a third of that area has also
been observed at 612 MHz. B\,1834+620 is situated in one of the early
regions of the WENSS, the so-called `Mini-Survey' region which has
been observed at both frequencies.  A total intensity radio map of
B\,1834+620 at 612 MHz is presented in Fig. \ref{fig:1834_wenss}a. Its
radio morphology resembles that of `four beads on a string'. The four
components are well aligned and have a quite similar brightness and
spacing.  Since the beam size of the WSRT at 325 MHz is twice as large
it does not reveal new details and therefore we do not show a radio
map of the source at this frequency.  The WENSS has also observed
linear polarization. Because of large, hour-angle dependent ionosperic
Faraday effects no reliable linear polarization could be measured at
325~MHz.

B\,1834+620 has also been observed during the NRAO VLA Sky Survey
(NVSS, Condon et al. 1998) which has been conducted at 1400~MHz at
comparable resolution to the WENSS. The NVSS supplies maps of the
Stokes' $Q$ and $U$ parameters, which we have used to map the linear
polarized intensity. See Fig. \ref{fig:1834_wenss}b for a 1.4-GHz
radio contourmap with polarized intensity vectors overplotted.  We
find that at this frequency the position angles of the observed
$E$-field polarization vectors differ by $\sim\!15\degr$ between the
northern and the southern half of the source.

\subsection{WSRT observations}

\subsubsection{1.4-GHz observations} 

We have observed B\,1834+620 at two frequencies with the WSRT. First,
we have obtained a short observation at 1.4~GHz, with 6 integrations
of 4 min. each at different hour-angles. The total bandwidth was
60~MHz and was centered at 1395~MHz.  The telescope gains were
calibrated to the flux density scale of Baars et al. (1977) using
3C\,286 as primary calibrator. The data have been processed and mapped
using the NFRA {\sc newstar} package. The non-uniform $(u,v)$-coverage
of these observations has resulted a highly elliptical
beam-shape. However, the minor axis of this beam is almost parallel to
the radio axis of the source, and so we are able to separate the four
main components. Linear polarization has been detected towards these,
with similar polarization angles as in the NVSS-data. The fractional
polarizations are 10\% (northern outer component), 21\% (northern
inner component), 20\% (southern inner component) and 13\% (southern
outer component).  We show no radio maps of these observations; as a
result of the sparse sampling of the $(u,v)$-plane the 1.4-GHz
observations reveal no new details.

\subsubsection{840-MHz observations}

To investigate the polarization at a frequency in between the WENSS
612-MHz and NVSS observations we have used the WSRT with the recently
installed UHF-High frontends at frequencies around 840~MHz
($\sim\!36$-cm wavelength).  We have integrated for 12 hr on March 31,
1998. Of the eight 10-MHz channels, four suffered from severe
interference and could not be used at all. The remaining four channels
are centered at 815, 845, 865 and 875 MHz. Gain and phase calibration
was obtained by observations of 3C\,48 and 3C\,286. We used the flux
density scale of Baars et al. (1977) for absolute gain calibration.
The data were calibrated and mapped using the NFRA {\sc newstar}
data-reduction package. The morphology of the source in the 840-MHz
map closely resembles that in the 612-MHz map shown in
Fig. \ref{fig:1834_wenss}a and therefore we do not show it.

\subsection{VLA observations}

\subsubsection{8.4-GHz observations}

We observed B\,1834+620 with the VLA in its D-configuration on
August 3, 1996. Two standard 50~MHz bands were used, centered at
8435~MHz and 8485~MHz. The total integration time was 25 minutes. The
telescope gains were calibrated on the Baars et al. (1977) scale using
3C\,147 and 3C\,286 as primary flux density calibrators. The source
J1845+401 was used as phase calibrator. The polarization angles have
been calibrated using 3C\,286 as reference source.

\begin{figure*}
\psfig{figure=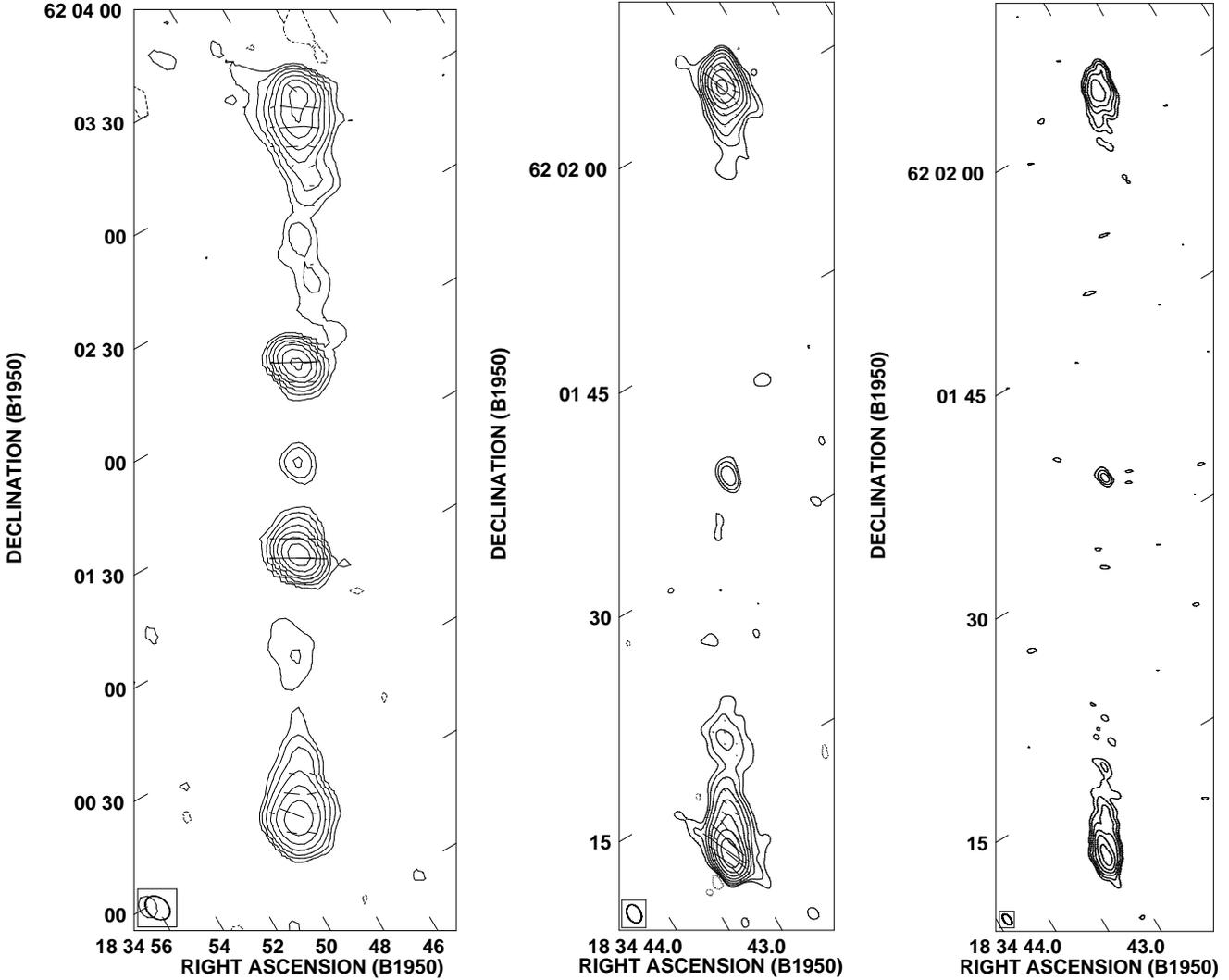,width=\textwidth}
\caption{\label{fig:1834_inner} {\bf a} (left) Contour plot of
B\,1834+620 from the 8.4-GHz VLA observations. The contours start at
0.22 mJy beam$^{-1}$. Subsequent contours are logarithmic with a
factor of 2 between levels. Dashed contours designate negative
levels. The vectors indicate the direction of the $E$-field of the
linearly polarized intensity. Their length corresponds to the
polarized intensity and is 0.25 mJy beam$^{-1}$ per arcsec. {\bf b}
(middle) Contour plot of the inner structure, only, from the 1.4 GHz
VLA observations. The contours start at 0.20 mJy beam$^{-1}$ and
increase logarithmic with a factor of 2 between
levels. Dashed contours designate negative levels. The vectors
indicate the direction of the $E$-field of the linearly polarized
intensity. Their length corresponds to the polarized intensity and is
2.5 mJy beam$^{-1}$ per arcsec. {\bf c} (right) Contour plot of the
inner structure, only, from the 5-GHz VLA observations. The contours
start at 0.21 mJy beam$^{-1}$ and increase logarithmic with a factor
of 2 between levels.}
\end{figure*}

The data have been reduced and mapped using the {\sc nrao aips}
software package. After initial {\sc clean}ing, several
self-calibrations were performed to improve the data-quality.  The
total intensity map, corrected for primary beam attenuation, is shown
in Fig. \ref{fig:1834_inner}a with $E$-field polarization vectors
overplotted on the total intensity contours.

A radio core is clearly detected in between two bright and only
slightly resolved inner components.  The two outer components have
been resolved and have a morphology similar to edge-brightened
FRII-type radio lobes.

\begin{figure*}
\psfig{figure=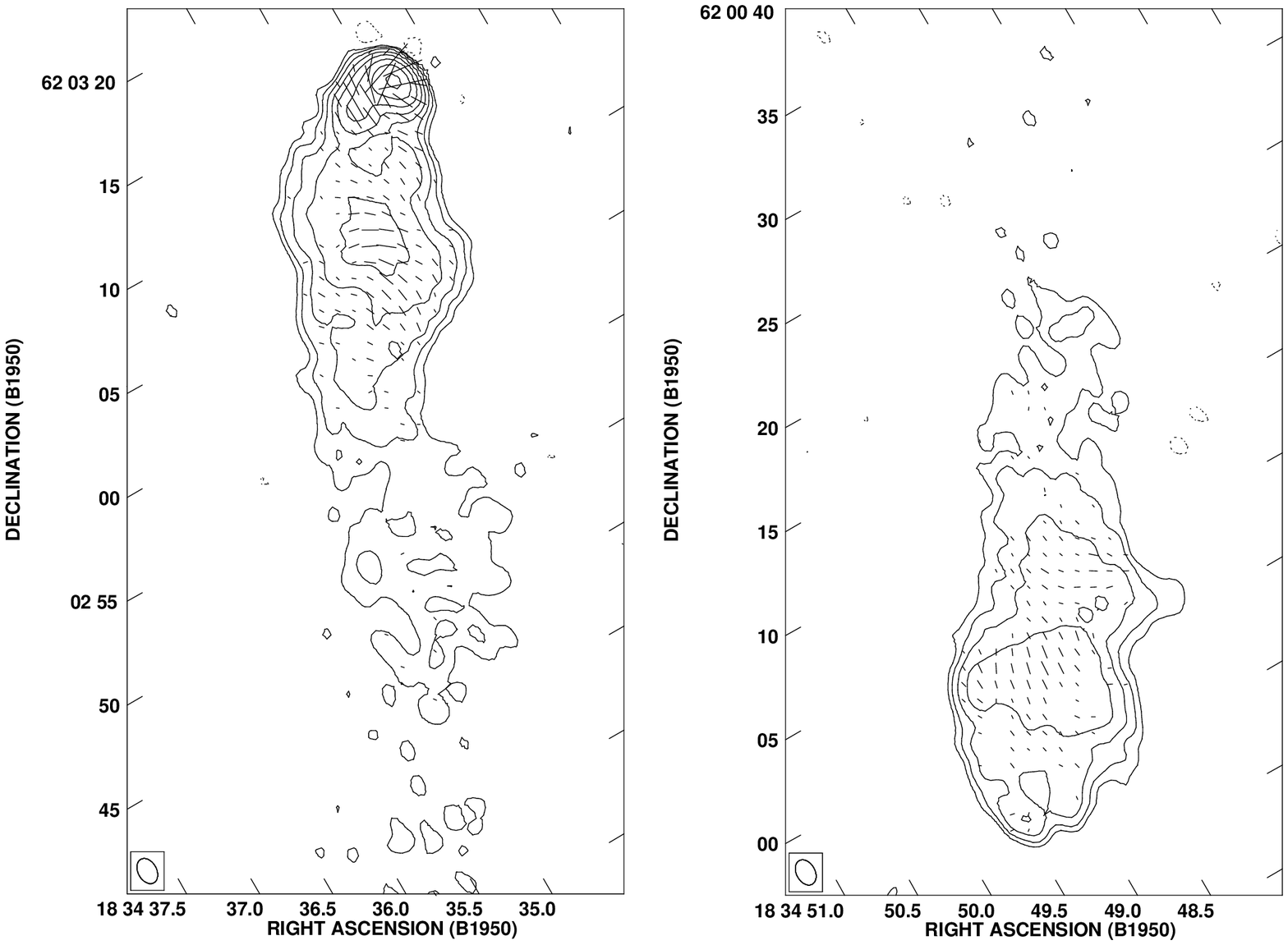,width=\textwidth}
\caption{\label{fig:1834_14outer} Contour plots from the 1.4-GHz VLA
observations of the two outer radio lobes. The contours start at 0.20
mJy beam$^{-1}$ and subsequent contours are logarithmic with a factor
of 2 between levels. Dashed contours designate negative levels. We
also plot the observed vectors of the $E$-field of the linearly
polarized intensity. Their length corresponds to the polarized
intensity and is 1.0 mJy beam$^{-1}$ per arcsec.}
\end{figure*}

\subsubsection{1.4-GHz observations}

To map of the inner components and the outer radio lobes in more
detail, we have used the VLA in its A-configuration at 1435 MHz. To
avoid bandwidth smearing at the outer edges of the source, we have
used the spectral line mode with 8 channels of 3.125~MHz bandwidth
each, resulting in a total bandwidth of 25~MHz. We further used the
full polarization mode to map the linear polarization at high
resolution.  The telescope gains were calibrated using the primary
calibrator 3C\,286 and set to the scale of Baars et al. (1977). For
phase calibration, we have used several observations of the source
J1845+401. The total on-source integration time was
$\sim\!55$~minutes. One of the channels had to be flagged because of
poor data quality.  The data of the remaining channels were then
combined into a single dataset. Total and polarized intensity maps
have been made after several passes of {\sc clean} and
self-calibration.

The inner components resemble FRII-type radio lobes (see
Fig. \ref{fig:1834_inner}b). The southern inner lobe somewhat
resembles the southern outer lobe, as observed at 8.4 GHz
(Fig. \ref{fig:1834_inner}a). The radio core is well
detected and is unresolved.  By fitting it with a Gaussian (using the
{\sc jmfit} program in {\sc aips}) we find that its B1950.0 position
is $18^h34^m41\fs03 \pm 0\fs02$ in Right Ascension and
$+62^{\circ}01\arcmin35\farcs3 \pm 0\farcs1$ in Declination.

The northern outer lobe (see Fig. \ref{fig:1834_14outer}a) has a
bright compact hotspot, which is not reproduced in the southern
lobe. At the resolution of these observations the ratio of peak flux
density of the northern to the southern hotspot is 19.3\,.

\subsubsection{5-GHz observations}

The inner structure has been mapped at 5~GHz with the VLA in its BnA
configuration on June 28, 1998. We observed B\,1834+620 for 20
minutes. The data were calibrated using 3C\,48 as primary flux density
calibrator. After initial mapping, several passes of phase
selfcalibration were made. The final map is somewhat degraded due to
the presence of the large outer lobes, which we were not able to map
accurately at the large bandwidth of these observations. The inner
structure has been mapped at almost twice the resolution of the
1.4-GHz VLA observations (see Fig. \ref{fig:1834_inner}c). The
southern inner lobe reveals a leading hotspot, while the northern
inner lobe remains rather featureless. The core remains unresolved.

\subsection{Identification of the host galaxy}
\label{sec:1834_identification}

We have examined an optical image from the Digitized Sky Survey (DSS)
to search for the host galaxy of B\,1834+620. However, at the position
of the radio core no optical counterpart could be identified.
Therefore, an optical CCD image has been taken with the 3-m Shane
telescope at Lick observatory on April 19th 1996. The images have been
taken using the direct imaging mode on the KAST spectrograph, equipped
with a Reticon CCD at a spatial scale of $0\,\farcs74$/pixel. A
R$_{\rm s}$-filter has been used, which is designed to suppress the
Sodium background illumination caused by urban areas (e.g. Djogorvski
1985).  Four exposures of 2 minutes each have been made, each shifted
by $\sim\!30\arcsec$~in either Right Ascension or Declination to allow
cosmic ray and bad pixel removal. The total integration time is
therefore 8 minutes.  For flux calibration the photometrical standard
star Feige~34 has been observed (Djorgovski 1985). The seeing during
these observations was $\sim 2\,\farcs3$. The astrometry has been
fixed using several stars in the field whose coordinates have been
measured on the DSS with an accuracy of $\sim\!1\arcsec$.  An overlay
of the optical image with the 1.4-GHz VLA observations (see
Fig. \ref{fig:1834_opt}) shows that the host galaxy is the brightest
member of an apparently compact group of three objects. A similar image
in shown by Lara et al. (1999). The R$_{\rm
s}$ magnitude of the host galaxy, integrated in a circular aperture with a radius of
$5\arcsec$ centered on the peak of maximum intensity, is $19.9 \pm 0.1$\,. The two
nearby galaxies have R$_{\rm s}$-magnitudes of $\sim\!20.4$ and
$\sim\!21.1$. Since their separation from the host galaxy is only a
few arcsecond (i.e. a few tens of kpc at the redshift of the source), 
it is very likely that they belong to the same group.

The amount of galactic extinction towards the host galaxy is small.
We have measured the H{\sc i} column density, $N({\rm H{\sc i}})$, in the
Leiden-Dwingeloo H{\sc i}-survey (Hartmann 1994). We find
$N({\rm H{\sc i}}) = 4.56\times10^{20}$~cm$^{-2}$. Using a conversion
factor of
$N({\rm H{\sc i}})/E(B-V)\!=\!5.6\times10^{21}$~cm$^{-2}$\,mag$^{-1}$
(Burstein \& Heiles 1978) we find $E(B-V)\!=\!0.08$, and therefore the
galactic absorption in the R$_{\rm s}$-band yields $A_{\rm
R_s}\!=\!0.19$ mag.  If we correct the magnitude of the host galaxy
for galactic extinction we thus find R$_{\rm s} = 19.7 \pm 0.1$ mag.

\subsection{Optical spectroscopy}

\begin{figure}
\psfig{figure=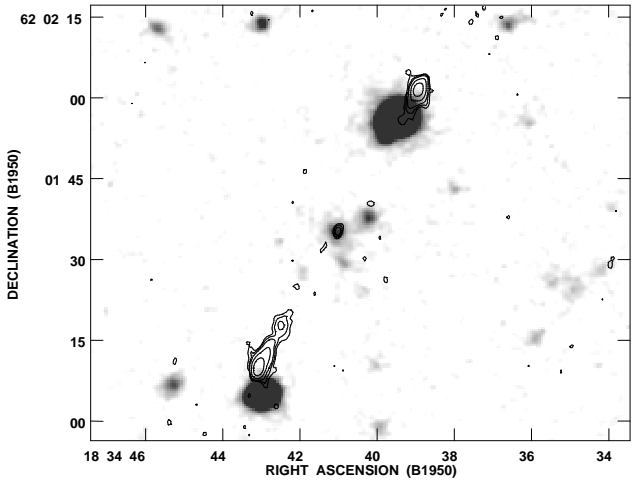,angle=270,width=\columnwidth}
\caption{\label{fig:1834_opt}Optical R$_{\rm s}$-band image of the
host galaxy of B\,1834+620, with the 1.4-GHz VLA radio
contours overlaid. The radio core has been identified with a R$_{\rm s} = 19.7
\pm 0.1$ mag galaxy, which is a member of a small and compact group
of three galaxies. The two bright objects near the inner radio lobes
have been spectroscopically identified as stars.}
\end{figure}

\begin{figure}
\psfig{figure=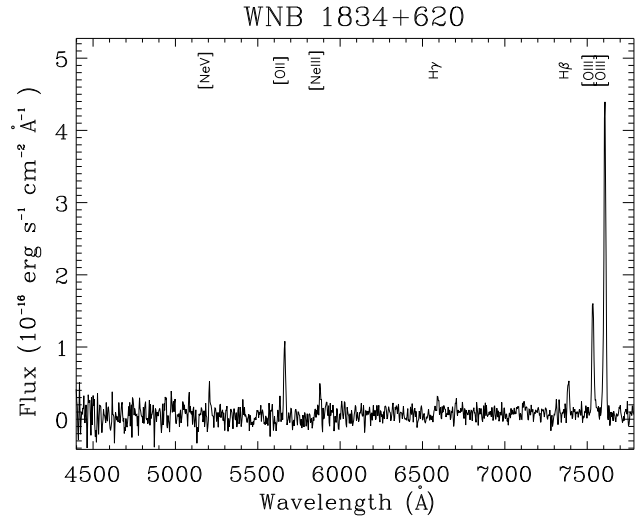,angle=90,width=\columnwidth}
\caption{\label{fig:1834_spectrum} Optical spectrum of the host galaxy
of B\,1834+620. The spectrum has been corrected for galactic
extinction, using $E(B-V) = 0.08$, and for atmospheric
absorption. Identified emission lines have been indicated.}
\end{figure}

To obtain a spectrum of the host galaxy we have used the 2.5-m INT
telescope on La Palma on April 8th, 1996. We have used the IDS
spectrograph with the R300V grating, equipped with a 1k$\!\times\!$1k
TEK chip. In this setup, one pixel on the CCD corresponds to 3.32\AA~
in the dispersion direction, and $0\,\farcs74$ in the spatial
direction. We used a central wavelength of 6000\AA~and a slitwidth of
$2\arcsec$ projected on the sky, yielding a resolution of
$\sim\!7.6$\AA.  The seeing during these observations was
$\sim\!1\,\farcs7$. We have aligned the slit over the two bright
starlike objects at R.A. $18^h 34^m 39\fs72$, Dec. $62\degr 01\arcmin
56\farcs3$ and R.A. $18^h 34^m 43\fs09$ Dec. $62\degr 01\arcmin
05\farcs5$ (B1950.0), since the host galaxy is situated exactly on the
line connecting these two stars (see Fig. \ref{fig:1834_opt}). Two
600-s exposures have been taken at a mean airmass of 1.25\,. These
have been averaged to obtain the final spectrum and to enable cosmic
ray removal.  Wavelength calibration was done by observations of
internal arc-lamps, and checked against several bright skylines. Flux
calibration was obtained by observations of several
spectrophotometrical standard stars during the night. The
datareduction was done using the {\sc longslit} package in the {\sc
nrao iraf} data reduction software.  We have extracted the spectrum of
the host galaxy and the two star-like objects in a $4\arcsec$-wide
spatial aperture.  The star-like objects indeed had stellar spectra,
ruling out any possible connection with the radio source.  The
spectrum of the host galaxy is dominated by the $[$O{\sc ii}$]$3727
and $[$O{\sc iii}$]$4959,5007 emission lines. The $[$O{\sc
iii}$]$5007-line lies in an atmospheric absorption band. We corrected
for this absorption by fitting the continuum of the northern bright
star by a low-order polynomial and by subsequently dividing the
spectrum of the galaxy by the normalized ratio of this fit and the
observed spectrumn of that star. We have corrected for galactic
extinction using an $E(B-V)$ of 0.08, as found in the previous
section. The resulting spectrum is shown in
Fig. \ref{fig:1834_spectrum}.

We have measured the positions and flux densities of the indicated
emission lines (see Tab. \ref{tab:linefluxes}).  We find a mean
redshift of the emission lines of $0.5194 \pm 0.0002$. This is somewhat
higher than the redshift of 0.518 quoted by Lara et al. (1999), who
used slighty lower resolution spectral data. 
From the measured flux densities we have calculated the
emitted powers, assuming isotropic emission. The results are
presented in Tab. \ref{tab:linefluxes}.  At a
redshift of 0.5194 the $2\arcsec \times 4\arcsec$ size of the aperture used in the extraction corresponds
to a physical aperture of $\sim\!15 \times 30$ kpc, and should thus
contain most of the optical galaxy. We find that the total
contribution of the emission lines to the measured R$_{\rm s}$-magnitude
of the host galaxy is $\la 20\%$, or $\la 0.25$ mag. The image of the
host galaxy is thus dominated by star light, and not by emission-line
gas.

\begin{table*}
\begin{minipage}{\textwidth}
\setlength{\tabcolsep}{12pt}
\caption{\label{tab:linefluxes} Properties of the emission lines in
our spectrum of the host galaxy of B\,1834+620. The fluxes have been
corrected for atmospheric absorption and for galactic extinction using
$E(B-V) = 0.08$ and the interstellar reddening curve of Cardelli et
al. (1989). The fluxes and peak positions have been measured using
Gaussian fits to the observed line profiles. The FWHM are deconvolved
and translated to velocity in the rest frame. Upper limits are
calculated assuming a width comparable to the resolution (7.6\AA). The
numbers between the brackets denote the errors in the measured
quantities. The powers have been calculated assuming isotropic
emission.}
\begin{tabular}{lccccc}
\hline
Line & $\lambda_{\rm obs}$ & $z$ & FWHM & \multicolumn{1}{c}{Flux} &  \multicolumn{1}{c}{Power} \\
 & $[$\AA$]$ & & \multicolumn{1}{c}{$[$km\,s$^{-1}]$} & \multicolumn{1}{c}{\small $[10^{-16}$~erg\,s$^{-1}$\,cm$^{-2}]$} & \multicolumn{1}{c}{\small $[10^{42}$~erg\,s$^{-1}]$}\\
\hline 
$[$Ne{\sc v}$]$3426 & 5206.6 & 0.5197 & $< 440$ & 3.80 (0.20) & 1.46 (0.08) \\
$[$O{\sc ii}$]$3727 & 5663.3 & 0.5195 & $< 400$ & 13.7 (0.3)\phantom{1} & 5.25 (0.11) \\
$[$Ne{\sc iii}$]$3869 & 5878.8 & 0.5195 & $< 390$ & 4.50 (0.27) & 1.72 (0.10) \\
H$\gamma$4340 & 6593.4 & 0.5192 & $< 350$ & 2.34 (0.19) & 0.90 (0.07) \\
H$\beta$4861 & 7385.9 & 0.5194 & $< 310$ & 6.29 (0.16) & 2.41 (0.06) \\
$[$O{\sc iii}$]$4959 & 7534.4 & 0.5193 & 294 & 23.7 (0.2)\phantom{1} & 9.08 (0.08) \\
$[$O{\sc iii}$]$5007~$^a$ & 7607.2 & 0.5193 & 283 & 63\phantom{.1} (13)\phantom{.1} & 24\phantom{.1} (5)\phantom{.11} \\
\hline
\end{tabular}
\ \\
\begin{flushleft}
{Notes:}\\ 
{$^a$~This line coincides with an atmospheric absorption
band for which we corrected using a stellar spectrum (see the text for
details). We estimate that the resulting values are accurate to
20\%.}\\
\end{flushleft}
\end{minipage}
\end{table*}

\section{Results}
\label{sec:results}

In this section we present and discuss the results of our radio
observations. We will discuss the morphology, flux densities, radio
spectrum and polarization properties of B\,1834+620.

\subsection{Radio morphology and linear size}

The 1.4 and 8.4-GHz VLA observations show that the outer components of
B\,1834+620 have a FRII-type radio morphology.  The northern outer
lobe possesses a bright compact hotspot at its head. The extended lobe
regions behind the head of the lobes are quite similar on both sides
of the source, in terms of morphology, surface brightness distribution
and polarization structure.

The most interesting features are the bright inner components. The 1.4
and 5-GHz VLA observations show that these are FRII-type radio
lobes. They are situated very symmetrically about the radio core. The
misalignment of the radio axis determined by the inner components (the
`inner' radio axis) with that determined by the outer components (the
`outer' radio axis) is only $\sim\!2\degr$.  The radio morphology of
the southern inner lobe in our 1.4-GHz VLA observations is remarkably
similar to that of the southern outer lobe as seen in the 8.4-GHz VLA
map (see Fig. \ref{fig:1834_inner}).

The total angular size of B\,1834+620 is 230 arcsec. At a redshift of
0.5194 this corresponds to a projected linear size of 1660~kpc.
B\,1834+620 is the third largest radio source known at $z > 0.5$,
after 8C\,0821+695 at $z = 0.538$ with a projected linear size of
3.0~Mpc (Lacy et al. 1993) and the $z=0.634$ quasar HE\,1127-1304
which has a projected linear size of 2.4 Mpc (Bhatnagar, Gopal-Krishna
\& Wisotzki 1998).

\subsection{Armlength asymmetries}

The overall morphology of B\,1834+620 is very symmetrical. To
measure the armlength-ratio of the outer lobes, we use the 1.4-GHz VLA
map. We find that the distance from the core to the hotspot in the
outer southern lobe is $4.0\arcsec$ larger than that to the hotspot in
the northern outer lobe ($116.9\arcsec$ versus $112.9\arcsec$). The
armlength ratio, defined here as the ratio of the separation between
the core and the farthest hotspot to that of the core and the nearest
hotspot, is $1.035\pm0.002$\,.  For the inner lobes we find an
armlength ratio of $1.050\pm 0.007$ with the northern lobe being the
more distant one. The inner structure is therefore slightly more
asymmetrical as the outer structure, and the asymmetry is opposite
in sense.

\begin{table*}
\begin{minipage}{\textwidth}
\setlength{\tabcolsep}{12pt}
\caption{\label{tab:fluxes}
Radio flux densities of B\,1834+620 and its components. For the
frequencies of 612 MHz to 1400 MHz, the `components' are defined as in
Fig. \ref{fig:1834_boxes}. At higher frequencies, they are the actual
lobes of the indicated source structure.  The lower flux density at
1395~MHz as compared to the 1400-MHz NVSS flux density is most likely
a result of the poor $(u,v)$-coverage of the WSRT observations. The
VLA observations at 1435 and 4850~MHz could not be used to measure the
flux densities of the outer lobes, due to a lack of
sensitivity for structures of such angular scales. 
Errors in the flux densities include calibration
errors, which have been estimated to be 10\% at 38 and 151 MHz, 5\% at
1395 MHz and 4850 MHz (GB6 flux density), and 2\% for all other
measurements and errors due to noise in the map.}
\begin{tabular}{r r@{$\,\pm\,$}l r@{$\,\pm\,$}l r@{$\,\pm\,$}l r@{$\,\pm\,$}l r@{$\,\pm\,$}l r@{$\,\pm\,$}l}
\hline 
Freq. & \multicolumn{2}{c}{Total} & \multicolumn{4}{c}{North} &  \multicolumn{4}{c}{South} & \multicolumn{2}{c}{Core}\\
 &\multicolumn{2}{c}{ } & \multicolumn{2}{c}{\small Inner} & \multicolumn{2}{c}{\small Outer} & \multicolumn{2}{c}{\small Inner} & \multicolumn{2}{c}{\small Outer} \\
$[$MHz$]$ & \multicolumn{2}{c}{$[$Jy$]$} & \multicolumn{2}{c}{$[$mJy$]$} & \multicolumn{2}{c}{$[$mJy$]$} & \multicolumn{2}{c}{$[$mJy$]$} & \multicolumn{2}{c}{$[$mJy$]$} & \multicolumn{2}{c}{$[$mJy$]$} \\  
\hline 
38$^a$   & 19.1 & 1.9  \\
151$^b$  & 5.6  & 0.6  \\
327$^c$  & 2.96 & 0.06 \\
612$^d$  & 1.67 & 0.04 & 309 & 7 & 573 & 12 & 422 & 9 & 357 & 8 \\
845$^e$  & 1.22 & 0.03 & 228 & 5 & 435 & 9  & 298 & 6 & 259 & 5 \\
1395$^f$ & 0.72 & 0.04 & 136 & 7 & 254 & 13 & 176 & 9 & 152 & 8 \\ 
1400$^g$ & 0.80 & 0.02 & 144 & 3 & 275 & 6 & 198 & 4 & 177 & 4 \\
1435$^h$ & \multicolumn{2}{c}{ } & 77.2 & 2.2 & \multicolumn{2}{c}{ } & 122.1 & 3.6 & \multicolumn{2}{c}{ } & 1.52 & 0.08\\
4850$^i$ & 0.24 & 0.02 & 30.9 & 1.0 & \multicolumn{2}{c}{ } & 51.0 & 1.2 & \multicolumn{2}{c}{ } & 1.70 & 0.10 \\
8460$^j$ & 0.14 & 0.01 & 19.4 & 0.4 & 53.0 & 1.1 & 29.1 & 0.6 & 32.3 & 0.7 & 1.04 & 0.08 \\
\hline 
\end{tabular}
\vskip 1mm
Notes: a - 8C survey (Rees 1990, catalogue revised by Hales et al. 1995);
b - 6C survey (Hales et al. 1991); c - WENSS 92cm; d - WENSS 49cm; e - Our
WSRT observation; f - Our WSRT observation; g - NVSS; h - Our VLA
observation; i - Total flux density: GB6 survey (Gregory et al. 1996, but
measured in the map), individual components: Our VLA observations; j - Our VLA observations.\\
\end{minipage}
\end{table*}     

\begin{figure}
\psfig{figure=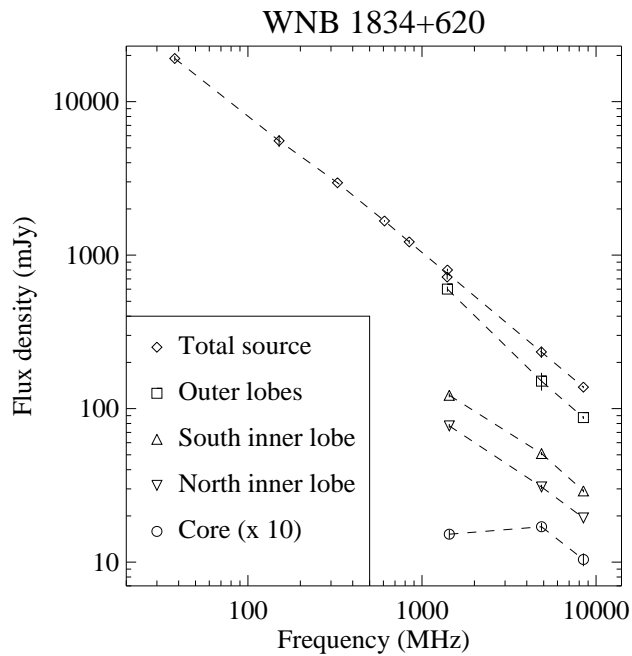,width=\columnwidth}
\caption{\label{fig:fluxes}The radio spectrum of B\,1834+620. We
have plotted the integrated flux density, the flux density of the
outer lobes calculated as the total flux minus the flux density of
the inner lobes and core, the flux densities of the inner lobes where
they could be measured from the radio maps, and the flux density of
the radio core multiplied by a factor of 10. The values
can be found in Tab. \ref{tab:fluxes}.}
\end{figure}

\subsection{Flux densities and spectral indices}
\label{sec:spectral_indices}

We have collected flux densities of B\,1834+620 at a variety of
frequencies, using our own radio data and data from the literature. The
results are presented in Tab. \ref{tab:fluxes}. For those components
that have been measured at similar resolution by Lara et al. (1999),
our results agree very well with theirs.. A plot of the radio
spectrum is given in Fig. \ref{fig:fluxes}, where we have differentiated
between various source components wherever the data allowed us to do so.
The spectrum of the radio core has a convex shape, resembling the spectrum of
Gigahertz Peaked Spectrum (GPS, e.g. Spoelstra et al. 1985) sources. Such
convex shapes at GHz-frequencies are often found for weak cores in radio
galaxies and probably indicates that they are very compact (e.g. Rudnick,
Jones \& Fiedler 1986).

The spectra of the inner components steepen between 5 and 8.4 GHz: The
spectral index of the southern inner component changes from
$\alpha_{1.4}^{5} = -0.72\pm0.05$ to $\alpha_5^{8.4} = -1.02 \pm
0.10$, that of the northern inner component from $\alpha_{1.4}^{5} =
-0.75 \pm 0.05$ to $\alpha_5^{8.4} = -0.84 \pm 0.10$. Between 1.4 and
5 GHz, the spectra of the inner components are flatter than that of
the source as a whole ($\alpha_{1.4}^{5} = -0.97 \pm 0.08$).  The
outer components are too heavily resolved to allow an accurate flux
density measurement in our 1.4 and 5-GHz VLA data. However, 
subtracting the flux density of the inner lobes and the core from the
integrated flux density yields the sum of the flux densities
of the two outer lobes.  The spectral index of
the outer components alone, determined by fitting these three flux
density measurements, is $\alpha_{1.4}^{8.4} = -1.07 \pm 0.02$.  The
inner components are barely resolved at 1.4 GHz; therefore we have not
mapped the spectral index distribution in these source components.

\begin{figure}
\psfig{figure=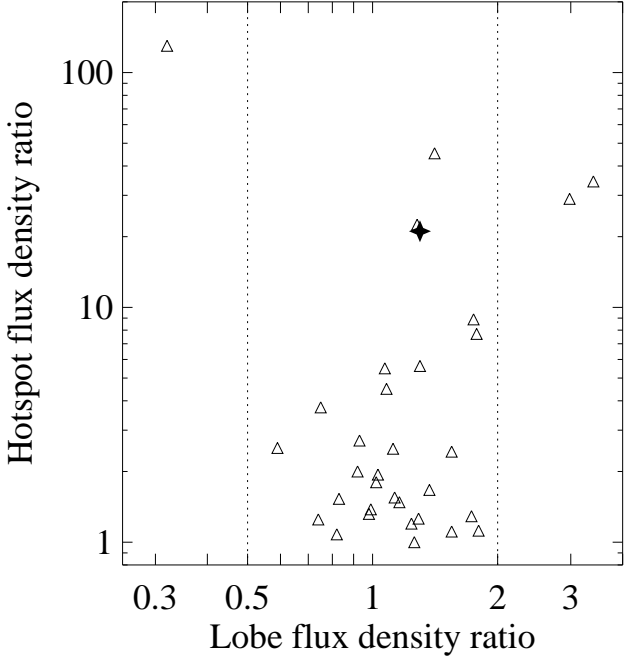,width=\columnwidth}
\caption{\label{fig:hotspots_lobes_flux_ratios} The hotspot flux
density ratio against the lobe flux density ratio of the $z<0.3$ 3CR
FRII-type sources of Hardcastle et al. (1998). A value smaller than
one indicates that the brighter hotspot is in the weaker radio lobe.
The flux densities are mostly measured at 8.4 GHz. The filled star
indicates the source B\,1834+620, measured at 1.4 GHz. The two
dashed lines delineate the region where the lobe flux densities are
within a factor of two of each other.}
\end{figure}

The flux density contrast between the hotspots in the outer two lobes
appears to be quite large (see Fig. \ref{fig:1834_14outer}). To
investigate how it compares to that of other radio sources, we have
used the data of Hardcastle et al. (1998). They present the flux
density at, mainly, 8.4 GHz of the primary (i.e. brightest) hotspot in
the two radio lobes of an almost complete sample of $z < 0.3$
FRII-type radio galaxies from the 3CR catalogue.  We define the flux
density ratio of the hotspots, $R_h$, in their sample by dividing the
flux density of the brightest hotspot by that of the weakest. We find
a median value of $R_h$ of $2.43$. Due to the generally small
asymmetry in hotspot spectral index within a single source, we expect
that $R_h$ will not be much different when measured at 1.4 GHz.  In
Fig. \ref{fig:hotspots_lobes_flux_ratios} we have plotted the hotspot
flux density ratio $R_h$ against the radio lobe flux density ratio,
$R_l$, defined as the ratio of the flux density of the lobe with
the brightest hotspot to that of the lobe with the weakest hotspot.
We subtracted the contribution from the hotspot in each lobe
first. The open triangles are sources from the sample of Hardcastle et
al. and are measured mostly at 8.4 GHz.  From our 1.4-GHz VLA
observations we measure a ratio of hotspot flux density in the outer
lobes of B\,1834+620 of $\sim\!20\,$. This has been indicated by the
filled star in the figure.  

We find that, in general, the asymmetries in the hotspot flux
densities are larger than those in the lobe flux densities. This
confirms earlier results obtained by, e.g., Macklin (1981). Further,
we find that the hotspot flux density in B\,1834+620 is among the
largest observed in FRII-type radio sources.
 
\begin{figure}
\psfig{figure=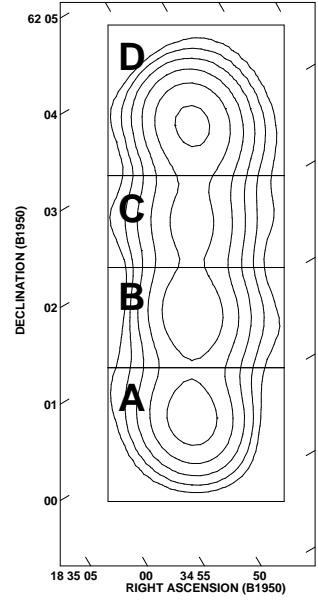,width=0.5\columnwidth}
\caption{\label{fig:1834_boxes}Contour plot of B\,1834+620 at 612 MHz,
after re-weighting the $(u,v)$-data from the WENSS with a Gaussian
taper of 1166\,m which results in a beamsize of
$51\arcsec\times45\arcsec$ (FWHM). The rectangular boxes, which have
been labeled A through D, indicate the four regions we have used in
the Rotation Measure analysis.}
\end{figure}

\begin{figure}
\psfig{figure=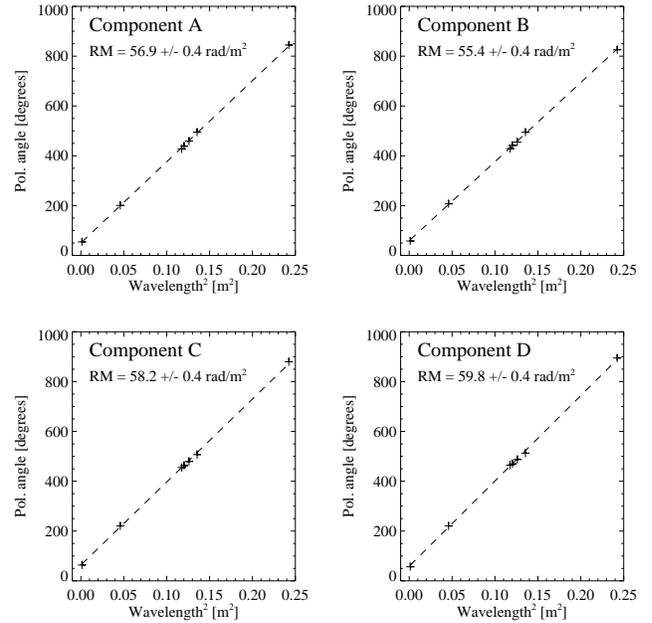,width=\columnwidth}
\caption{\label{fig:1834_rm}Plot of the polarization angles and the
best fit rotation measures of the four components of
B\,1834+620. The sizes of the symbols are not representative
of the errors in the individual measurements, which are always less
than $10\degr$.}
\end{figure}

\subsection{Depolarization and Rotation Measures}

We find that the degree of polarization of the inner lobes is
independent of the used beam-size (see Tab. \ref{tab:1834_fracpol}).
Both the 1.4-GHz VLA observations (beamsize $\sim\!1\farcs5$ FWHM) and
the WSRT (beamsize $\sim\!11\arcsec$ FWHM) measure equal fractional
polarizations, despite the fact the the VLA observations have resolved
the inner structure and the WSRT observations not. This implies that
the observed polarization structure is quite uniform, and that any
external Faraday rotation must be occuring on angular scales
comparable to or larger than the WSRT beam. Also, we find no evidence
for significant depolarization between 8.4 and 1.4 GHz towards the
inner lobes. This agrees with the results by Lara et al. (1999).

Using the WENSS 612-MHz, the WSRT 840-MHz, the NVSS 1.4-GHz and the
VLA 8.4-GHz observations we have determined the rotation measures (RM)
towards the four main components of B\,1834+620. We have mapped the
612 and 840-MHz WSRT data using a Gaussian $(u,v)$-taper of 1166 and
841m, respectively, to match them with the NVSS data.  The VLA 8.4-GHz
maps have been convolved to the NVSS resolution. Since artefacts may
arise from this we have given this data little weight in determining
the rotation measure.

We have defined four rectangular areas (see Fig. \ref{fig:1834_boxes})
in which we have integrated the Stokes' $Q$ and $U$ flux at each
frequency. The 840-MHz WSRT observations have four separate channels,
so that for each rectangular area we have box-averaged polarization angles
at 7 different frequencies.  The RMs have been determined by a 
least-squares linear fit of the polarization angles to $\lambda^2$,
where $\lambda$ is the wavelength of the observation. The data points
and their best fits are shown in Fig. \ref{fig:1834_rm}.

The found RMs roughly range from $+55$ to $+60$ rad\,m$^{-2}$. These
numbers agree with the lack of a detectable change in polarization
angle of the inner structures between the different channels of our
1.4-GHz VLA data. We note that the large RM-values ($276 \pm 7$ rad
m$^{-2}$) found by Lara et al. (1999) must be wrong. If the RM were
that large, we would have seen a significant change in the polarization
angle between the different channels in our 1.4 GHz observations ($\sim
3\degr$ per channel, or 25\degr over the whole range).
We would also have suffered from severe bandwidth depolarization in the
612-MHz data. Further, the change in the polarization angle measured
between the separate channels in the 840-MHz data are not reconcilable
with such a large value of RM. We therefore believe are our values are
the correct ones. If this is indeed true, the polarization angles in
the 5 GHz data of Lara et al. (1999) are wrong by almost exactly
90\degr. 

The small differences in RM that we measure vbetween the different
components suggest
that the major contribution to the rotation 
is galactic in origin, and not intrinsic
to the radio source. The low amount of variation of the RM over the
source, and the lack of depolarization towards the inner structure at
1.4 GHz, suggest the presence a low density environment around the
radio source. The largest RM contrast is that between components B and
D, and is only $4.4\!\pm\!0.8$ rad\,m$^{-2}$.

\begin{table}
\caption{\label{tab:1834_fracpol}
Fractional polarization of the inner lobes.}
\begin{tabular}{r c r@{$\,\pm\,$}l r@{$\,\pm\,$}l}
\hline
Freq & Instr. & \multicolumn{2}{c}{North} & \multicolumn{2}{c}{South}\\
$[$MHz$]$ & &\multicolumn{2}{c}{$[\%]$} & \multicolumn{2}{c}{$[\%]$}\\
\hline \\
1395 & WSRT & 20.7 & 0.8 & 19.9 & 0.8 \\
1435 & VLA  & 21.9 & 0.7 & 20.7 & 0.7 \\
8460 & VLA  & 19.1 & 0.7 & 18.7 & 0.7 \\
\hline \\
\end{tabular}
\end{table}

\section{Discussion}
\label{sec:discussion}

We will discuss the results of the observations assuming that the
inner structures have formed after an interruption of the jet
formation in the AGN, as discussed in Papers I \& II. We will
investigate how the emission line properties relate to the observed
radio properties, and we will estimate the time scale of the
interruption. Also, we will estimate the age of the inner structures,
their advance velocities and the density in their ambient medium.

\subsection{Emission-line luminosity and line-ratios}

Tadhunter et al. (1998) have investigated the optical to radio
correlations of a complete sample of powerful radio galaxies in the
redshift range between 0.1 and 0.7\,. Their results agree well with
earlier results, in that there is a significant correlation between
the radio power and the emission-line luminosity of these radio
sources.  Since the optical spectrum of B\,1834+620 shows strong
emission lines, one would expect that the associated radio source is
quite powerful.  If we assume that the AGN is currently only feeding
the inner radio structure, we have to compare the optical properties
of the AGN with the radio properties of the inner structure alone.

Following Tadhunter et al., we have calculated the total radio power,
$P_{rad}$, of the inner structure and find $P^{inner}_{rad}\!=\!
10^{43.5}$ erg\,s$^{-1}$. Comparing the luminosity of the optical
emission lines (see Tab. \ref{tab:linefluxes}) and the radio power of
the inner structure of B\,1834+620 with those of the sources in
Tadhunter et al., we find that the radio power is an order of
magnitude too low to account for the observed emission-line
luminosity.  This is also the case when we compare the $[$O{\sc
iii}$]$5007/$[$O{\sc ii}$]$3727 line ratio with the radio power. On
the other hand, when we compare the $[$O{\sc iii}$]$5007/$[$O{\sc
ii}$]$3727 line ratio with the emission-line luminosity of $[$O{\sc
iii}$]$5007, $[$O{\sc ii}$]$3727 and H$\beta$, we find no
discrepancies between B\,1834+620 and the sources presented in
Tadhunter et al.  It indicates that the physical conditions, such as
the density and ionization state, in the narrow-line region of the
host galaxy of B\,1834+620 must be quite similar to those in more
radio-powerful narrow-line radio galaxies. This, and the high
emission-line luminosity, strongly suggests that the energy output of
the AGN is comparable to that of radio sources which are an order of
magnitude more radio-luminous. In case of B\,1834+620 this 
hints towards a low efficiency for the transformation of jet energy into
synchrotron radiation. A low density of the environment of the inner
lobes may account for this.

\subsection{The time scale of the interruption of the jet}

Under the assumption that the inner structures have formed after an
interruption of the jet formation in the AGN we can constrain the
length of time of the interruption in B\,1834+620.  Since hotspots
fade rapidly (within $10^4 - 10^5$ yr, depending on their size; e.g. Paper
II) the presence of a bright hotspot in the northern outer lobe
implies that it still receives jet material. The travel time of jet
material from the core to the northern hotspot, $t_j$, is
$D_{hs}/v_j$, with $D_{hs}$ the physical distance from the core to the
hotspot and $v_j$ the velocity of the material flowing down the jet;
$v_j$ is probably close the speed of light, $c$, and we will therefore
assume it is equal to it. Further, we will assume that the
interruption is instantaneous and that the last of the jet material
also flows at a velocity of $c$ through the jet. This yields $t_j =
(2.66/\sin\theta)$ Myr, where $\theta$ is the angle between the line
of sight and line connecting the core to the northern hotspot. The
outer structure of B\,1834+620 has a projected linear size of 1660
kpc. This makes it unlikely that the orientation of the radio axis is
close to the line of sight. The absence of broad emission lines, a
nuclear non-thermal continuum in the optical spectrum or a dominating
unresolved QSO in our optical image further indicates that the angle
between the radio axis and the plane of the sky is $\la 45\degr$
(cf. Barthel 1989). This results in a travel time $t_j$ between 2.66
and 3.76 Myr.

\begin{figure}
\psfig{figure=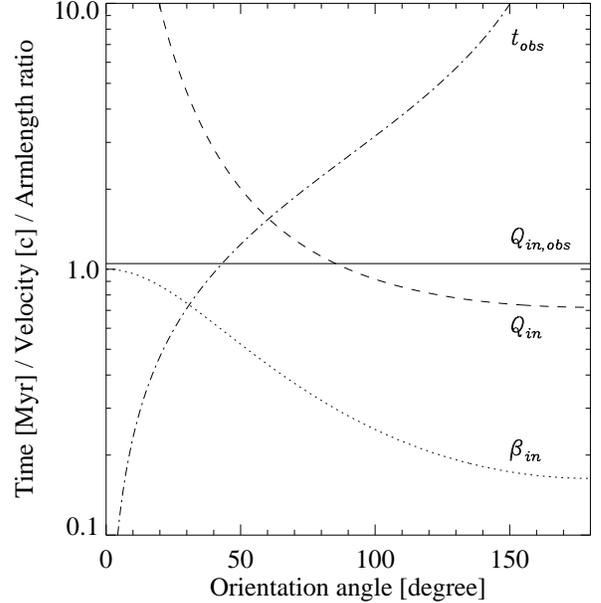,width=\columnwidth}
\caption{\label{fig:1834_velocity}Three source parameters of the inner
lobes of B\,1834+620 as a function of the orientation angle $\theta$
between the line of sight and the northern radio axis. The dot-dashed
line gives the observed time difference, $t_{obs}$, between the
ejection of the last jet material from the core and the fading of the
northern hotspot in Myr.  The dotted line gives the true advance
velocity, $\beta_{in}$, of the northern inner component in units of
$c$ using the presence of the northern hotpsot to limit the available
time. The dashed line gives the expected armlength ratio, $Q_{in}$, of
the inner source assuming equal advance velocities
$\beta_{in}(\theta)$ of both inner radio lobes.  The solid line is the
observed armlength ratio, $Q_{in, obs}$.  The intersection of the
solid line and the dashed line gives an estimate of the orientation of
the inner source under the assumption of an equal advance velocity for
both lobes, and is $85.3\degr \pm 1.5\degr$ (see the text for
details).}
\end{figure}

However, because of light travel time effects the orientation of the
source complicates matters. If the northern outer lobe is oriented
away from us (i.e. $\theta\! > \!  90\degr$), it will take longer for
the information that the last jet material has arrived in the hotspot
to reach the observer than if it is oriented towards us. The {\sl
observed} time difference, $t_{obs}$, between the ejection of the last
jet material from the core and its arrival in the outer hotspot is
given by $t_{obs} = (D_{hs}/c) (1 - \cos\theta) = (2.66/\sin\theta) (1-\cos\theta)$ Myr. We have plotted $t_{obs}$ as a function of
$\theta$ in Fig. \ref{fig:1834_velocity} (the dot-dashed line). If the
jet is interrupted for a time scale which surpasses $t_{obs}$, the
inner structure and the hotspot can not be observed simultaneously.
Therefore, the interruption of the jet activity must have lasted less
than between 1.1 and 6.4 Myr, depending on the orientation of the
radio source. Further, within this length of time, the inner structure
must have developed so that even less time is available for the
interruption itself.

\subsection{The age and velocity of the inner structure}
\label{sec:ages}

We will assume that the restart of the jet after it has stopped is
instantaneous, since this provides a lower limit to the velocity of
the inner structure.  From the distance between the core and the
northern inner lobe, $D_{in,obs}$ we can calculate its apparent
(i.e. projected) advance velocity $\beta_{in,app} =
D_{in,obs}/(c\,t_{obs})$, where $\beta_{in,app}$ is in units of
$c$. The true advance velocity of the inner lobe, $\beta_{in}$, can be
found in a way which is equivalent to finding true velocities for
superluminal VLBI sources. We find $\beta_{in} = \beta_{in,app} /
(\sin\theta + \beta_{in,app} \cos\theta)$.  The projected distance
between the core and the northern inner component is 229 kpc or
$7.47\!\times\!10^5$ lightyears.  Therefore, $\beta_{in,app} =(0.281
\sin\theta)/(1-\cos\theta)$.  The true advance velocity $\beta_{in}$
is shown as a function of $\theta$ in Fig. \ref{fig:1834_velocity}.

The allowed advance velocities of the inner structure range between
$\sim\!0.16c$, if $\theta$ is close to $180\degr$, and $1c$, if
$\theta$ is close to $0\degr$. For reasonable values of $\theta$,
i.e. $45\degr \!\le\! \theta \!\le\! 135\degr$, we find an advance velocity of
the inner structure between $0.19c$ and $0.57c$, which translates into
an age between 1.9 and 5.8 Myr.  

If we assume that the two inner components are advancing equally fast, 
this should lead to asymmetries in
the observed armlength ratio. For the inner structure of B\,1834+620
we measure an armlength ratio, $Q_{in, obs}$, of only $1.050 \pm
0.007$. Using $Q_{in} = (1+\beta_{in}\cos\theta)/(1-\beta_{in}\cos\theta)$, we have plotted In
Fig. \ref{fig:1834_velocity} the expected armlength-ratio as a
function of $\theta$, using $\beta_{in}(\theta)$ as the advance
velocity. A value of $Q_{in}$ smaller than 1 indicates that the
northern inner lobe is observed closer to the nucleus than the
southern inner lobe.  We find that under the assumptions we have made,
$\theta = 85.3\degr \pm 1.5\degr$. The advance velocity of the inner
structures is then $0.298c \pm 0.006c$, and their age is
$2.51 \pm 0.05$ Myr.

A second method to constrain the advance velocity of the inner lobes
uses the observed large asymmetry in hotspot flux density in
B\,1834+620 (see Fig. \ref{fig:hotspots_lobes_flux_ratios}). We will
assume that the hotspot in the southern outer lobe is weak because no
jet material is arriving there anymore, i.e. it has faded.  This need
not be the case of the observed asymmetry. Other possibilities include
variations of the jet power, instabilities in the jet and
inhomogeneities in the external medium.  The projected distance from
the core to the southern outer hotspot is 844 kpc and that to the
northern outer hotspot 815 kpc. Therefore, assuming that the
termination of the jet production occurs simultaneously on both sides
and that jet material travels at the speed of light, the difference in
travel-time for the jet material between the two hotspots is
$0.096/\sin\theta$ Myr. For a distant observer, the light travel-time
difference between the northern and the southern hotspot is
$5.412/\tan\theta$ Myr. Assuming that the northern hotspot is
currently still active, implies that the light travel-time difference
must be larger than the difference in travel-time of the jet material.
This leads to the constraint that $\theta \ga 91\degr$, i.e. the
northern side of the source must be pointing away from us. For this
orientation, the upper limit of the advance velocity of the northern
inner lobe is $0.277c$ and the corresponding minimal age is $2.70$
Myr. This agrees reasonably with the value determined using the
armlength asymmetry. We stress again that all these velocity estimates
assume a direct restart of the interrupted jet.

\subsection{The density of the environment of the inner structure}

Even the lower limit of the advance velocity of the hotspots of
$0.19c$ is quite high, when compared to such values in other radio
galaxies (e.g. Alexander \& Leahy 1987). In general, the highest
advance velocities are found in the most powerful radio galaxies, but
the inner structure of B\,1834+620 is only of modest luminosity.
This strongly suggests that it must be expanding in an extremely low
density environment. We estimate the density of the environment of the
inner lobes, $\rho_a$, following the method outlined in Daly (1995)
and Wellman, Daly \& Wan (1997). This assumes that the lobe
propagation is ram pressure controlled, so that $p_l = 0.75 \rho_a
v^2_l$, with $v_l$ the propagation velocity of the radio lobe and
$p_l$ the non-thermal pressure in the radio lobe, which can be
estimated from the minimum magnetic field strength $B_{min}$. In the
inner lobes of B\,1834+620 we find $B_{min} = 16.5\mu$G. We note
that this is probably an overestimate, since we use the integrated
flux of the lobes and this is dominated by the hotspots. Wellman et
al. (1997) find that the true magnetic field strength in radio lobes
is probably smaller than $B_{min}$ by a factor of $\sim\!4$. When we
take this into account, we find a non-thermal pressure in the inner
lobes of $3.8\times10^{-11}$ dyn\,cm$^{-2}$.  Adopting the lower limit
on the lobe advance velocity of $0.19c$, which we found earlier, the
ambient density $\rho_a = 1.6\times10^{-30}$ gr\,cm$^{-3}$. Assuming
that the ambient medium of the inner lobes is comparable to an IGM
with a mean particle mass of 1.4 amu, we find an ambient particle
density $n_a$ of $8\times10^{-7}$ cm$^{-3}$.  Note that this value
scales with the inner lobe advance velocity as $v_l^{-2}$ and
therefore strictly is an upper limit. Wellman et al. (1997) find $n_a
\sim10^{-4} - 10^{-3}$ cm$^{-3}$ for a sample of powerful FRII-type
radio sources. This is almost three orders of magnitude above our
value. Part of this may be related to powerful FRII-type sources being
situated in much denser cluster environments, whereas large sources
such as B\,1834+620 probably are not. However, this cannot explain a
contrast of three orders of magnitude.

In the restarted radio jet scenario (Paper II) the new radio jet is
propagating in the old cocoon formed by the first phase of
activity. The density inside the cocoon is expected to be much less
than the density of the ambient IGM. In Paper II we predict a particle
density of $4.1\times10^{-6}$ cm$^{-3}$ inside the old cocoon. This is
a factor of five above our estimate, which even is an upper
limit. Given the high number of assumptions in both our method to find
the ambient density and in the model used in Paper II, we do not think
this is distressing.

\section{Summary and Conclusions}
\label{sec:conclusions}
We have investigated the radio structure of the double-double radio
galaxy B\,1834+620, using a wide variety of frequencies and
resolutions. Both the outer and the inner radio lobes have an
FRII-type morphology.  The northern outer lobe has a bright hotspot,
in contrast to the southern outer lobe. The brightness-ratio is
$\sim\!20$, which is rather extreme when compared to values obtained
for a nearly complete sample of $z<0.3$ FRII-type radio galaxies from
Hardcastle et al. (1998).

We have obtained an optical spectrum and an ${\rm R_s}$-band CCD image
of the host galaxy of this peculiar radio source.  We find that it is
the brightest member of a group of three galaxies. Its ${\rm
R_s}$-magnitude, corrected for galactic extinction, is
$19.7\pm0.1$. The spectrum shows prominent emission lines for which we
measure a redshift of $0.5194\pm 0.0002$.

As we have argued in Papers I and II, the radio properties of the
outer and inner source strongly point towards interrupted jet activity
as the cause of the DDRG-strucure.  The presence of the hotspot in the
northern lobe then limits the length of time of the interruption to
the range between 1.1 and 6.8 Myr, depending on the orientation of the
source.  On geometrical grounds we have estimated the advance velocity
of the inner lobes to be in the range of $0.19c - 0.57c$, which
corresponds to an age between 1.9 and 5.8 Myr. If we assume that the
armlength asymmetry of the inner structure is the result of light
travel time effects, the advance velocity of the inner structures must
be $0.298c \pm 0.006c$, and their (observed) age $2.51 \pm 0.05$ Myr.
Similarly, if the southern hotspot has disappeared because of an end
to its energy supply, we can constrain the orientation angle between
the line of sight and the line connecting the core to the northern
hotspot to $> 91\degr$ and set an the upper limit to the velocity of
the northern inner lobe of $0.277c$. All these results indicate large
advance velocities of the inner structures, much larger than measured
in other FRII-type radio sources on size-scales of a few hundred kpc. 
We estimate
the density in the medium surrounding the inner lobes to be $\la
1.6\times10^{-30}$ gr\,cm$^{-3}$ (particle density $\la
8\times10^{-7}$ cm$^{-3}$), which is a factor of five below the
density predicted in Paper II for this source.  An extremely low
density of the ambient medium of the inner structure is further
supported by the contrast between the relatively high luminosity of the
optical emission lines, and the low radio power of the inner lobes.

The results presented here strongly suggest that the inner lobes are
advancing rapidly in a low density environment, in agreement with the
model presented in 
Paper II. The numbers we find for, e.g., the density of the gas
surrounding the inner lobes or the age of the inner structure, do not completely agree with the model
predictions, but the difference are always within a factor of two to
five. Given the extreme conditions inside the cocoon predicted by the
model, we find this not alarming.  

\section*{Acknowledgments}

The INT is operated on the island of La Palma by the Isaac Newton
Group in the Spanish Observatorio del Roque de los Muchachos of the
Instituto de Astrofisica de Canarias.  The Westerbork Synthesis Radio
Telescope (WSRT) is operated by the Netherlands Foundation for
Research in Astronomy (NFRA) with financial support of the Netherlands
Organization for Scientific Research (NWO).  The National Radio
Astronomy Observatory (NRAO) is operated by Associated Universities,
Inc., and is a facility of the National Science Foundation (NSF).
This research has made use of the NASA/IPAC Extragalactic Database
(NED) which is operated by the Jet Propulsion Laboratory, California
Institute of Technology, under contract with the National Aeronautics
and Space Administration.  We would like to thank B. Clark for
scheduling the 8.4~GHz VLA observations and C. de Breuck and W. van
Breugel for the Lick observations. H. Sanghera and
D. Dallacasa are thanked for their help in the early stages of this
project. C. Kaiser, P. Best and M. Lehnert are thanked for many 
useful discussions. We also thank the referee, J. P. Leahy, for his comments
on the manuscript.

{}


\begin{thebibliography}{}
\bibitem{} Alexander, P., Leahy, J. P., 1987, MNRAS, 225, 1
\bibitem{} Baars, J. W. M., Genzel, R., Pauliny-Toth, I. I. K., Witzel, A., 1977, A\&A, 61, 99 
\bibitem{} Barthel, P. D., 1989, ApJ, 336, 606
\bibitem{} Bhatnagar, S., Gopal-Krishna, Wisotzki, L., 1998, MNRAS 299, L25
\bibitem{} Blandford, R., Rees, M., 1974, MNRAS, 169, 395
\bibitem{} Burstein, D., Heiles, C., 1978, ApJ, 225, 40
\bibitem{} Cardelli, J. A, Clayton, G. C., Mathis, J. S., 1989, ApJ, 345, 245
\bibitem{} Condon, J. J., Cotton, W. D., Greisen, E. W., Yin, Q. F., Perley, R. A., Taylor, G. B., Broderick, J. J., 1998, AJ, 115, 1693 
\bibitem{} Daly, R. A., 1994, ApJ, 426, 38
\bibitem{} Djorgovski, S., 1985, PASP, 97, 1119
\bibitem{} Fanaroff, B. L., Riley, J. M., 1974, MNRAS, 167, 31
\bibitem{} Gregory, P. C., Scott, W. K., Douglas, K., Condon, J. J., 1996, ApJS, 103, 427 
\bibitem{} Hales, S. E. G., Mayer, C. J., Warner, P. J., Baldwin, J. E., 1991, MNRAS, 251, 46 
\bibitem{} Hales, S. E. G., Waldram, E. M., Rees, N., Warner, P. J., 1995, MNRAS, 274, 447 
\bibitem{} Hardcastle, M. J., Alexander, P., Pooley, G. G., Riley, J. M., 1998, MNRAS, 296, 445
\bibitem{} Hartmann, D., 1994, PhD.-thesis, University of Leiden
\bibitem{} Kaiser, C. R., Schoenmakers, A. P., R\"{o}ttgering, H. J. A., 1999, MNRAS, accepted $[$Paper II$]$
\bibitem{} Lacy, M., Rawlings, S., Saunders, R., Warner, P. J., 1993, MNRAS, 264, 721 
\bibitem{} Macklin, J. T., 1981, MNRAS, 196, 967
\bibitem{} Rees, N., 1990, MNRAS, 244, 233
\bibitem{} Rengelink, R., Tang, Y., de Bruyn, A. G., Miley, G. K., Bremer, M. N., R\"{o}ttgering, H. J. A., Bremer, M. A. R., 1997, A\&AS, 124, 259  
\bibitem{} Rudnick, L., Jones, T. W., Fiedler, R., 1986, AJ, 91, 1011
\bibitem{} Scheuer, P. A. G., 1974, MNRAS, 166, 513
\bibitem{} Schoenmakers, A. P., de Bruyn, A. G., R\"{o}ttgering, H. J. A., van der Laan, H., Kaiser, C. R., 1999, MNRAS, accepted $[$Paper I$]$
\bibitem{} Spoelstra, T. A. T., Patnaik, A. R., Gopal-Krishna, 1985, A\&A, 152,
38
\bibitem{} Tadhunter, C. N., Morganti, R., Robinson, A., Dickson, R., Villar-Martin, M., Fosbury, R. A. E., 1998, MNRAS, 298, 1035 
\bibitem{} Wellman, G. F., Daly, R. A., Wan, L., 1997, ApJ, 480, 96

\end{thebibliography}
\end{document}